\documentclass[times, review, 10pt]{elsarticle}

\usepackage{amsmath, amssymb} %, amsfonts, amscd, mathtools, physics}
\usepackage{orcidlink}
\usepackage{changepage}

\DeclareUnicodeCharacter{2061}{}

\journal{Pattern Recognition}

\begin{document}

\begin{frontmatter}

\title{\textit{sentropy}: A Python Package for Revealing Hidden Differences in Complex Datasets}

\author[1]{Phuc~Nguyen,~PhD\orcidlink{0000-0001-9993-8434}}

\author[2]{Rohit~Arora,~PhD\orcidlink{0000-0001-7128-6403}}

\author[3]{Elliot~D.~Hill,~MSc}

\author[1]{Jasper~Braun,~PhD\orcidlink{0000-0003-1250-4399}}

\author[1]{Alexandra~Morgan\orcidlink{0000-0001-7787-0547}}

\author[1]{Liza~M.~Quintana,~MD\orcidlink{0000-0002-5043-7425}}

\author[4]{Gabrielle~Mazzoni}

\author[5]{Ghee~Rye~Lee,~MMSc\orcidlink{0000-0001-6614-0223}}

\author[6]{Rima~Arnaout,~MD\orcidlink{0000-0002-7134-0040}\fnref{fn1}}

\author[1]{Ramy~Arnaout,~MD, DPhil\orcidlink{0000-0001-6955-9310}\fnref{fn1}\corref{cor1}}
\ead{rarnaout@bidmc.harvard.edu}

\affiliation[1]{organization={Department of Pathology, Beth Israel Deaconess Medical Center},
            city={Boston},
            postcode={02115}, 
            state={MA},
            country={USA}}

\affiliation[2]{organization={Novo Nordisk},
            city={Plainsboro},
            postcode={08536}, 
            state={NJ},
            country={USA}}

\affiliation[3]{organization={Duke University},
            city={Durham},
            postcode={27708}, 
            state={NC},
            country={USA}}
            
\affiliation[4]{organization={University of Virginia},
            city={Charlottesville},
            postcode={22903}, 
            state={VA},
            country={USA}}

\affiliation[5]{organization={Ohio State College of Medicine, The Ohio State University},
            city={Columbus},
            postcode={43210}, 
            state={OH},
            country={USA}}

\affiliation[6]{organization={Department of Medicine, the Bakar Computational Health Sciences Institute, and the UCSF UC Berkeley Joint Program for Computational Precision Health, University of California San Francisco},
            city={San Francisco},
            postcode={94143}, 
            state={CA},
            country={USA}}

\cortext[cor1]{Corresponding author}

\fntext[fn1]{This work was supported by the Gordon and Betty Moore Foundation, the Food and Drug Administration, and the NIH under grants R01HL150394, R01HL150394-SI, R01AI148747, and R01AI148747-SI.}

\begin{abstract}
Machine-learning datasets are typically characterized by measuring their size and class balance. However, there exists a richer and potentially more useful set of measures, termed S-entropy (similarity-sensitive entropy), that incorporate elements' frequencies and between-element similarities. Although these have been available in the R and Julia programming languages for other applications, they have not been as readily available in Python, which is widely used for machine learning, and are not easily applied to machine-learning-sized datasets without special coding considerations. To address these issues, we developed \textit{sentropy}, a Python package that calculates S-entropy and is tailored to large datasets. \textit{sentropy} can calculate any of the frequency-sensitive measures of Hill's D-number framework and their similarity-sensitive counterparts. \textit{sentropy} also outputs measures that compare datasets. We first briefly review S-entropy, illustrating how it incorporates elements' frequencies and elements' pairwise similarities. We then describe \textit{sentropy}'s key features and usage. We end with several examples—immunomics, metagenomics, computational pathology, and medical imaging—illustrating \textit{sentropy}’s applicability across a range of dataset types and fields.
\end{abstract}

\bigskip
\begin{keyword}
data science \sep diversity \sep frequency \sep similarity \sep Shannon entropy \sep Simpson's index \sep metagenomics \sep immunomics \sep computational pathology \sep medical imaging \sep machine learning \sep Python
\end{keyword}

\end{frontmatter}

% \tableofcontents

\section{Introduction}

Assessing the size and quality of a dataset is an important part of machine learning in many fields (Table \ref{table:1}). How many unique elements are there, and how do they relate to each other? How do they cluster? Summary statistics provide initial answers and can form the basis for deeper questions. Historically, different statistics have featured more prominently in different areas of investigation. Examples include species richness in ecology, Shannon entropy in computer science, and Simpson's index in biology. In several fields, the practice is to bin elements before calculating statistics. For example, in metagenomics, sequences are usually binned into operational taxonomic units (OTUs); in immunomics, antibody and T-cell receptor genes are often binned by lineage or antigen specificity. In machine learning, elements are binned by label to calculate class balance.

\begin{table*}[tb]
\begin{center}
\includegraphics[width=0.95\textwidth]{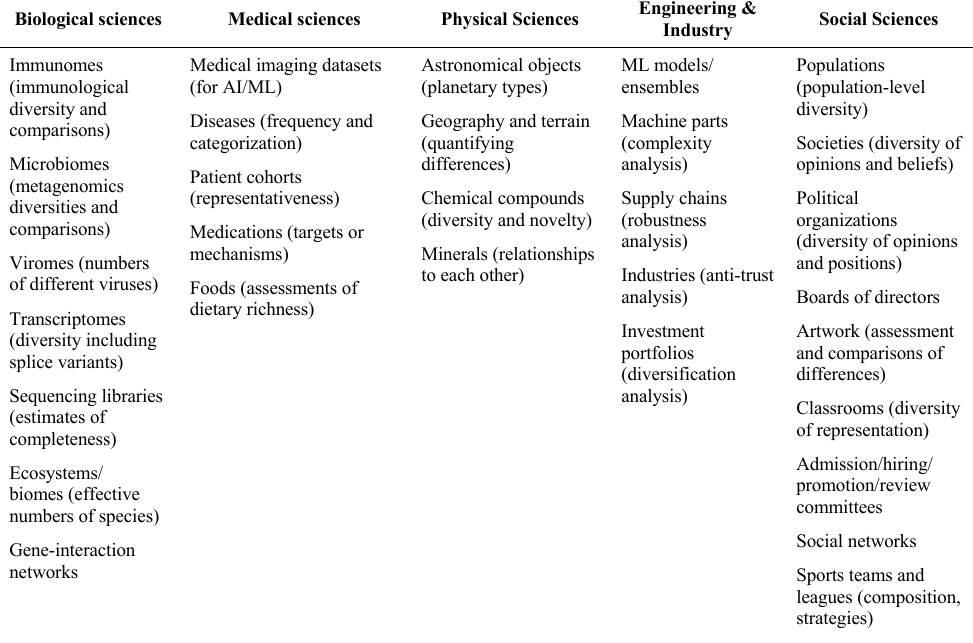}
\end{center}
\caption{
Areas and examples amenable to dataset composition analysis using \textit{sentropy}.
}
\label{table:1}
\end{table*}

\subsection{Frequency-sensitive diversity measures}

It has long been known that the summary statistics above, among others, are related by how they account for differences in the relative frequency of repeated dataset elements \cite{1}. For example, species richness, Shannon entropy, and Simpson's index, in that order, place an increasing emphasis on elements' frequencies, essentially up-weighting elements that appear more frequently by having them contribute more to the statistic’s final value. For example, in metagenomic datasets, species richness weights all OTUs the same regardless of how frequent they are, whereas in Simpson’s index common OTUs count more. In fact, these statistics are highly related to each other in that they are specific instances of a more general formula, known as Hill’s diversity framework, in which the key parameter is the frequency down-weighting q (sometimes known as the “viewpoint parameter”) \cite{2}:

\begin{equation}\label{eq:1}
D_q=(\sum _{i=1}^{n}p_{i}^{q})^{1/(1-q)}
\end{equation}

(Note $D_q$ also appears in the literature as ($^qD$.) In this equation, the dataset contains $n$ unique elements and $p_i$ is the frequency of the $i^{\text{th}}$ unique element in the dataset. In the limit of $q=1$, Eq. \ref{eq:1} approaches  $\exp{-\sum_i^n p_i \ln p_i}$, which the exponential of the Shannon entropy \cite{3}. The only substantive difference between species richness, Shannon entropy, Simpson's index, and many other summary statistics is in the value of $q$: $q=0$ for species richness (all elements count the same), $q=1$ for Shannon entropy (some down-weighting of rare elements), $q=2$ for Simpson's index (more down-weighting), and so on up to $q=\infty$ for the Berger-Parker index (maximum down-weighting) \cite{4}, for which the frequency of only the most common element matters.

\subsection{Effective numbers}

\begin{figure*}[t]
\centering
\includegraphics[width=.8\textwidth]{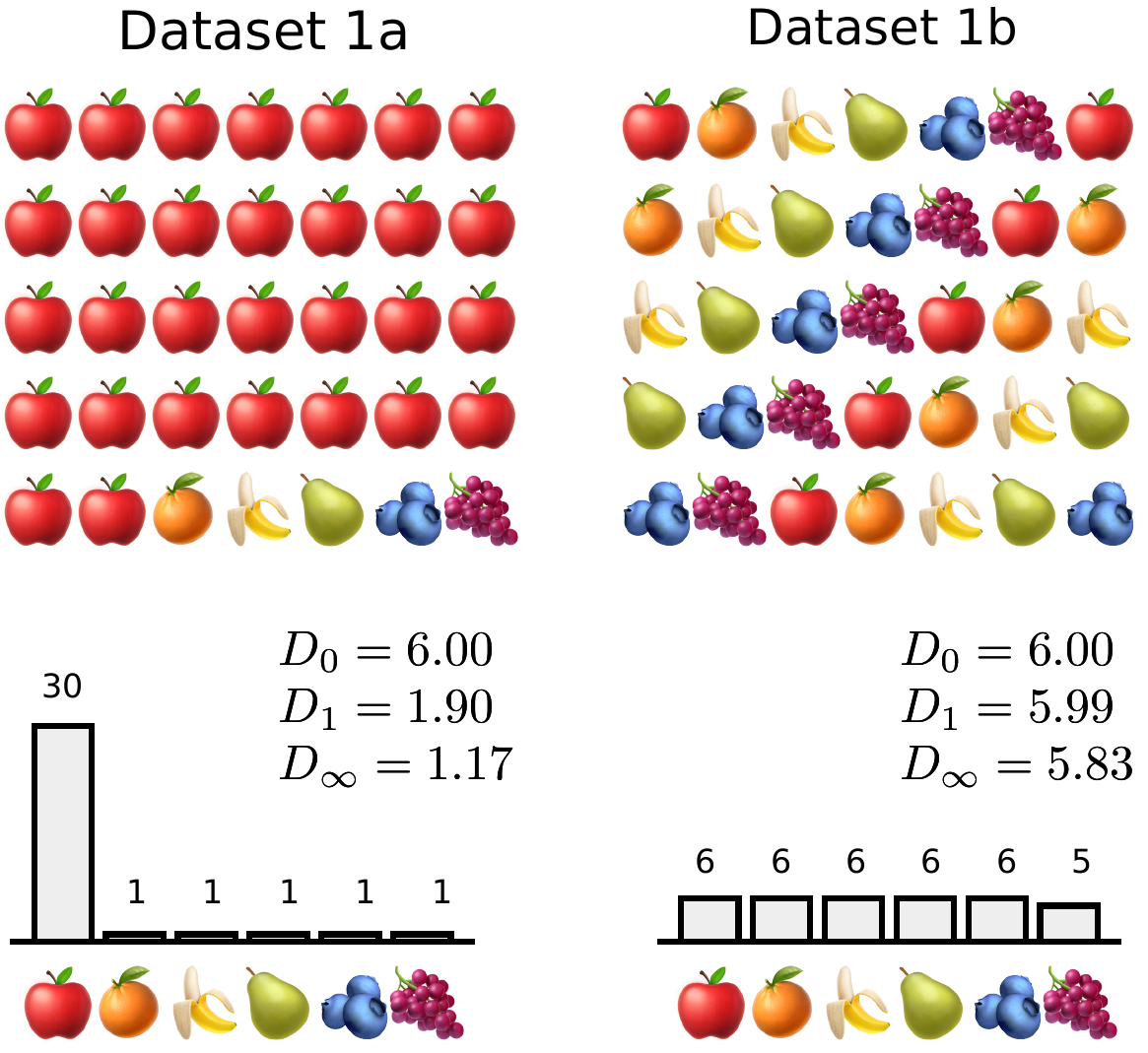}
\caption{
How frequency affects diversity (entropy). Each dataset contains six unique elements, so $D_0=6$ for each dataset (the ``0'' in $D_0$ makes $D_0$ frequency insensitive). Thus, if we ignore frequency, the two datasets are equally diverse. However, Dataset 1a is mostly apples, while Dataset 1b has a nearly uniform distribution of different fruits. Thus, Dataset 1b is intuitively more diverse. Consistent with intuition, $D_1$ is only 1.9 for Dataset 1a but nearly 6 for Dataset 2: for $q=1$, Dataset 1a effectively has only 1.9 unique elements: one can think of this as the apple counting as one unique element and the other fruits in Dataset 1a collectively counting as nearly (0.9) the equivalent of a second unique element. Up-weighting common elements can also be thought of discounting rare ones; thus, different values of $q$ can be thought of as different discount factors, and therefore lead to different effective numbers. For example, for $q=\infty$, $D_\infty=1.17$ for Dataset 1a and 5.83 for Dataset 2: with maximal discounting, Dataset 1a effectively has only fractionally more than a single element (effectively nearly all apples), whereas Dataset 1b still has close to all six (with a slight discount because there are only five bunches of grapes vs. six of each of the other fruits). The \textit{sentropy} package can calculate any frequency-sensitive diversity measure.
}
\label{fig:1}
\end{figure*}

The D numbers of Hill's framework yield the so-called effective number forms of Shannon entropy, Simpson's index, and so on. These give the effective number of elements in a dataset, for a given weight $q$. For example, if two immunology datasets contain the same number of unique antibody sequences but the first dataset has a sequence that accounts for 90\% of all sequences in the dataset (as can happen in leukemia), the first dataset will have a lower effective number of sequences for $q>0$. Likely the effective number of species will be not much more than 1, since to a good approximation---``effectively''---the dataset consists of just the one dominant sequence. Fig. \ref{fig:1} illustrates this concept in a discipline-agnostic manner. The commonly used forms of the above statistics are simple mathematical transformations of the effective-number forms: for example, Shannon entropy is the logarithm of $D_1$, while Simpson’s index is the reciprocal of $D_2$. For the investigator, the advantages of using effective numbers are, first, effective numbers all have the same units, allowing them to be easily compared; and second, they behave nicely as datasets are combined or split apart. Non-effective-number forms generally lack these properties \cite{5}. 

The effective number of elements in a dataset (at some $q$) is a natural measure of dataset size, complexity, or diversity \cite{6}. Because this number can also be thought of as a measure of how diverse the dataset is, the D numbers are also known as diversity measures (or diversity indices): ``diversities of order q.'' More specifically, when $q>0$, they are frequency-sensitive diversity measures. Within- and between-dataset measures are known as alpha and beta diversity, respectively.

We note the intimate connection between these measures and the Rényi entropies (of which Shannon entropy is the best known): the D numbers are simply the exponentials of the Rényi entropies. For this reason, it is useful to think of diversity and entropy interchangeably, as the same fundamental concept, just expressed in different units: effective numbers for diversity and bits or nats for entropy. Effective numbers are preferred for their several advantages \cite{6}.

\subsection{Similarity-sensitive diversity measures}

When grouping elements is called for, such as for OTUs and antibody/ TCR clusters, a similar approach to Eq. \ref{eq:1} can be used, provided that the similarity between each pair of elements is accounted for. Such accounting is possible through an extension of Hill’s framework that outputs similarity-sensitive versions of the D numbers, as shown by Leinster and Cobbold \cite{2}. We refer to this as similarity-sensitive diversity, similarity-sensitive entropy, or S-entropy for short and will express them in effective-number form. For each pair of unique elements in the dataset, the investigator supplies a similarity between 0 and 1, with 0 meaning the two elements are completely different and 1 meaning they are identical. Similarities are provided as entries in a table or matrix. This is called the similarity matrix, $Z$, a square matrix of dimension $n$ (where again $n$ is the number of unique elements). In the context of network science or graph theory, $Z$ can be thought of as an adjacency matrix with self-edges added to make the diagonal entries all equal to 1.

In practice, especially for large datasets, $Z$ is populated computationally according to an investigator-provided formula or rule. Binning can be understood as a special case where the cell $Z_{ij}$ that corresponds to the similarity between elements $i$ and $j$ is set to 1 if $i$ and $j$ are in the same bin and set to 0 otherwise. For example, for OTUs the similarity is be set to 1 if two 16S rRNAs have greater than a threshold-level of sequence identity (e.g. $\geq95\%$ or $\geq99\%$), placing them in the same OTU, and 0 otherwise (but see below for a continuous alternative). However, in general similarities $Z_{ij}$ can take any value between 0 and 1. They can be fractional (0.9, 0.327..., etc.) or even asymmetric, with element $i$ more similar to element $j$ than element $j$ to element $i$; the decision is up to the investigator \cite{7}. Some examples of similarity measures include the well-known structural similarity index measure (SSIM) for images \cite{8}, \cite{9} and relative dissociation constants for proteins and other molecules \cite{10}.

Mathematically, incorporating Z yields the following equation:
\begin{equation}\label{eq:2}
    D_q^Z=(\sum_{i=1}^n p_i \mathbf{Zp}_i^{q-1})^{1/(1-q)}
\end{equation}

\begin{figure}[t]
\centering
\includegraphics[width=0.8\textwidth]{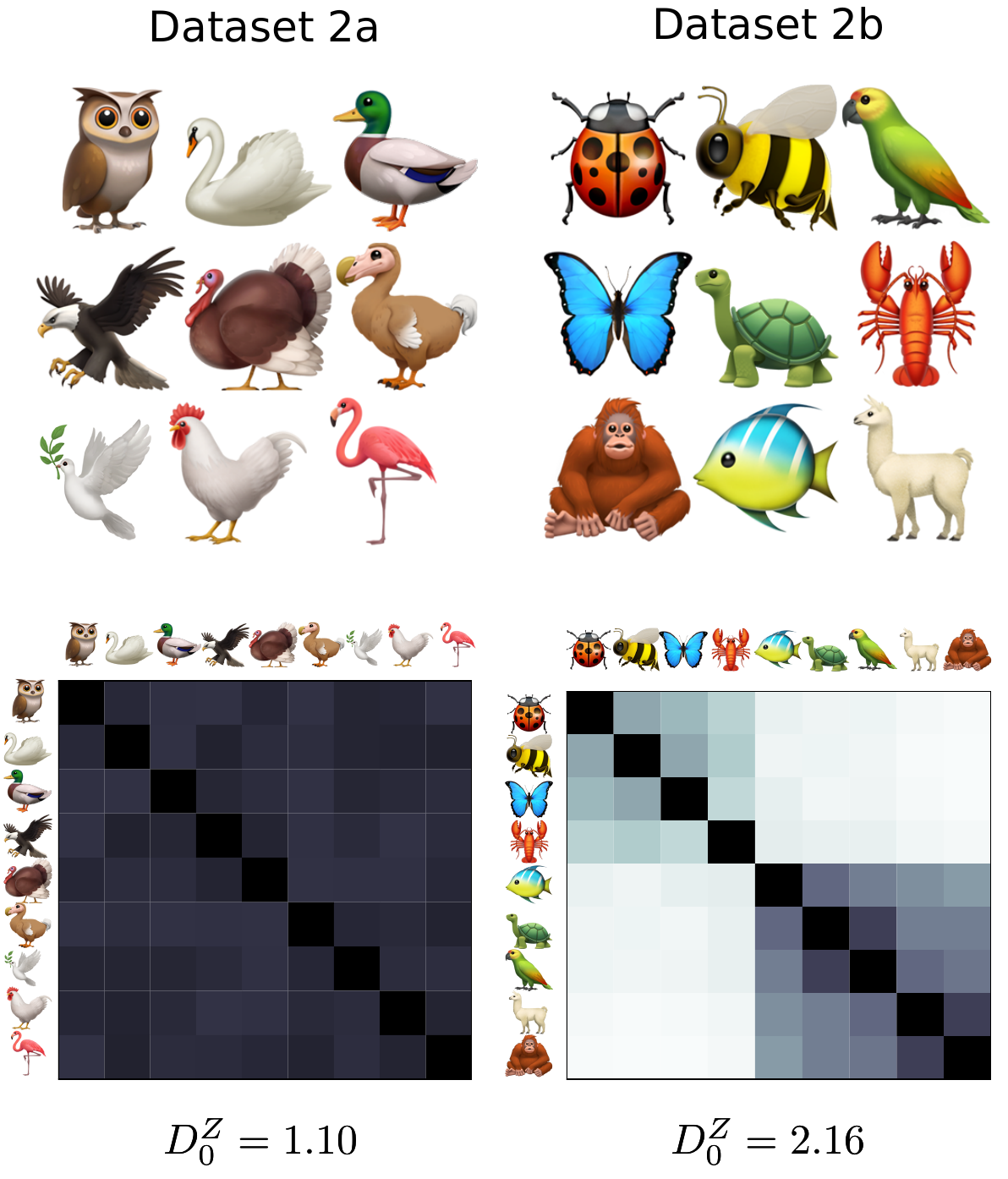}
\caption{
How similarity affects diversity. Each dataset contains nine species (top), each at equal frequencies, so $D_0=D_1=D_2=\dots=D_\infty=9$ for each dataset. Thus, the two datasets are equally diverse using similarity insensitive measures. However, Dataset 2a is all birds, whereas Dataset 2b contains a wider variety of animals, making Dataset 2b intuitively more diverse. The similarity matrices $Z$ are accordingly quite different (bottom). Consistent with the intuition, $D_0^Z$ is only 1.06 for Dataset 2a, but 2.16 for Dataset 2b: Dataset 2a effectively has only 1.06 species (further interpreted as it contains just a single class or type—here, birds; the ``extra'' 0.06 reflects intragroup diversity among bird species). Meanwhile, Dataset 2b effectively has 2.16 species, corresponding roughly to the number of taxonomic phyla represented (Arthropoda and Chordata, which form two visible clusters in the similarity matrix). The \textit{sentropy} package can calculate any similarity-sensitive diversity measure for any user-supplied definition of similarity.
}
\label{fig:2}
\end{figure}

Here, $\mathbf{Zp}_i=\sum_jZ_{ij}p_j$ is the ``ordinariness'' of the $i^\text{th}$ unique element, i.e. the cumulative similarity of the $i^\text{th}$ unique element to all the other elements of the dataset \cite{2}. The $Z$ in $D_q^Z$ designates these measures as being similarity-sensitive D numbers; the presence of $q$ in Eq. \ref{eq:2} means that these S-entropy measures are also frequency sensitive, like the $D_q$ numbers of \ref{eq:1}. Fig. \ref{fig:2} provides a discipline-agnostic illustration. Note that interpreting $Z$ as an adjacency matrix, Eq. \ref{eq:2} can be used to provide measures of network complexity in e.g. social networks or gene-interaction networks (Table \ref{table:1}).

\subsection{Advantages of similarity-sensitive measures}

S-entropic measures have several advantages over their similarity-insensitive counterparts. First, similarity-sensitive D numbers align better with an intuitive sense of what ``diversity'' and ``complexity'' should mean \cite{6}, \cite{7}. For example, given two otherwise identical 16S rRNA microbiome datasets, the dataset that represents the greater number of taxa is intuitively more diverse. Consistent with this intuition, it will have the higher $D_q^Z$ (see Metagenomics below). In contrast, the two datasets will have identical $D_q$, thus, the similarity-insensitive $D_q$ measures will miss the main difference between these datasets. 

Second, $D_q^Z$ measures capture the effects of grouping without the need to choose a binning threshold (e.g. 95\% vs. 99\% sequence identity for OTUs \cite{11}, \cite{12}, or a radius of 1 vs. 2 vs. 3 nucleotide differences for clone membership \cite{13}). There is also no need to place an element exclusively into a single group: multiple and/or fuzzy group membership is inherently allowed. This flexibility is especially useful for image datasets, where $D_q^Z$ simultaneously captures the number of classes as well as the diversity within and between classes \cite{14}.

Third, $D_q^Z$ measures provide rich mechanisms for exploring higher- and lower-level dataset structure, including a well-developed framework for partitioning within- (alpha) vs. between- (beta) dataset diversity. It also offers a ready method for investigating substructure using the ordinariness, $\mathbf{Zp}_i$. For example, to measure how ``Enterobacterales-like'' a given 16S rRNA dataset is, if the ordinariness to a given Enterobacterales sequence (or set of sequences) is high, the dataset is very Enterobacterales-like. This method could be useful for quantitative definitions and assessments of enterotypes (see Results).

We stress that the usefulness of S-entropy extends well beyond any one field. Table 1 lists several possible applications of S-entropic measures in different areas, from engineering to the natural and social sciences.

\subsection{Representativeness as a measure of overlap between datasets}

The S-entropic framework in Eq. 2 extends naturally to measures of between-dataset diversity, also known as beta diversity \cite{15}. Given a pair of datasets A and B,  $\bar{\beta}_A$ describes how distinctive dataset A is, with a value of 1 meaning indistinguishable and 2 meaning completely distinct. (We write out ``beta'' when referring to beta diversity so as not to confuse it with $\bar{\beta}$; unfortunately dual-use of this Greek letter is a convention in the literature.) Alternatively, we can use the representativeness, which is the inverse of distinctiveness and which is a more straightforward measure of overlap: representativeness of 1 means the dataset A is completely representative of the pair, whereas a zero-overlap dataset has 0.5 representativeness, because it is precisely half of the pair. (The minimum value is $1/N$ for $N$ datasets.) Mathematically, representativeness is written $\bar{\rho}$ and is the reciprocal of $\bar{\beta}$. We can also use (as an overlap measure) the diversity index $R$, which is the average of the representativenesses of the two datasets and which is known as the redundancy of the pair. Note, the original formulation of the framework \cite{2} uses the terms ``community'' and ``metacommunity'' in place of ``class''/``subset'' and ``overall''/``dataset,'' reflecting its roots in ecology; we prefer the latter in the \textit{sentropy} package.

\subsection{Prior work and present contribution}

The need for summary statistics that address the above-described issues has been well documented in metagenomics and immunology \cite{10}, \cite{11}, \cite{12}, \cite{16}. Other mathematical frameworks similar to the one in Eq. \ref{eq:2} have been proposed \cite{17}. Packages for calculating frequency- and similarity-sensitive $\alpha$ and $\beta$ diversities in effective-number form exist for the R and Julia programming languages \cite{18}, \cite{19}, but not to our knowledge for Python, which is more widely used, especially in machine learning.

Of note, large datasets present a special challenge for measuring similarity-sensitive diversity: because Eqs. \ref{eq:1}-\ref{eq:2} require calculating the similarity between each pair of the $n$ unique species in the dataset, and because these pairwise similarities are usually inputted as elements of an $n\times n$ matrix, direct implementations risk running out of computer memory when $n$ is large. For example, assuming standard 8-byte floating point precision, a single dataset of a million unique elements would require 8TB of memory. Immunomes, microbiomes, and imaging datasets routinely contain tens or hundreds of thousands of unique elements \cite{20}, as do transcriptomes, cell atlases, and other complex datasets. Moreover, a single study can involve tens to hundreds of such datasets (e.g., one per sample/volunteer/patient) and thereby many millions of unique elements in all. The R package \textit{rdiversity} and the Julia package \textit{Diversity} both require the similarity matrix to be stored in memory, but for the applications above, such matrices are likely to be too large to store in working memory and may even be too large to write to disk. For this reason, \textit{ sentropy} instead also implements on-the-fly row-by-row calculations, to better handle such cases. The Python package \textit{cdiversity} calculates Hill-number diversity but without similarity for immune repertoires \cite{22}.

In addition to the standard diversity indices in the $\beta$ family, we introduce two new quantities, $\hat{\rho}$ and $\hat{\beta}$ (``$\rho$-hat'' and ``$\beta$-hat''), which are normalized versions of $\bar{\rho}$ and $\bar{\beta}$. The reason is that $\bar{\rho}$ itself can be a very large number if there are a large number of datasets in the study. So, roughly speaking we can divide $\bar{\rho}$ by the number of subsets or classes to obtain a measure of redundancy that usually ranges from 0 to 1. (Note the upper bound may not be respected in extreme or pathologic cases, but these will be the exception.)

Furthermore, our package computes similarity-sensitive analogs of the Kullback-Leibler divergence (also known as the relative entropy) as well as of its R\'{e}nyi generalizations. We define the relative S-entropy between 2 abundance vectors $P$ and $Q$ to be:
\begin{equation}
D_{KL}^{Z}{(P||Q)} = \sum_{i \in \mathrm{supp}(P)} P_{i} \log{\left( \frac{(ZP)_{i}}{(ZQ)_{i}} \right)}
\end{equation}
In other words, starting with the usual definition of the relative entropy, we replace the $i^\text{th}$ component of $P$ and $Q$ inside the logarithm by the ordinariness vectors. Similarly, the similarity-sensitive analog of R\'{e}nyi entropies away from viewpoint parameter $1$ is:
\begin{equation}
D^{Z}_{q}{(P||Q)} = \frac{1}{q-1} \log{\left(\sum_{i \in \mathrm{supp}{(P)}} P_{i} \left( \frac{(ZP)_{i}}{(ZQ)_{i}} \right)^{q-1} \right)}
\end{equation}
The two formulas above are in entropy (i.e. traditional) forms. The corresponding D-number form are simply the exponential of the traditional forms.

\section{Methods}

\subsection{Python package} The package was developed and described as in the main text.

\subsection{Immunomes} Immunomes were obtained and processed, and similarities calculated, as previously described \cite{10}.

\subsection{Microbiomes} To construct each synthetic dataset, gaussian sampling to single-decimal-place precision was performed around each of a set of mean values between 0 and 100 to create a gaussian distribution of sequence frequencies around each mean. In Fig. \ref{fig:4}, the same number of sequences was sampled around each mean. In Fig. \ref{fig:5}, we sampled four times as many sequences around one of the means, which was selected as the dominant taxonomic unit defining the enterotype (a different dominant mean for each enterotype).

\subsection{Medical imaging} Images were selected at random from the dataset introduced and processed as previously described \cite{36}.

\subsubsection{Microscopy/Computational pathology} The 89,996-image PathMNIST training set is a component of MedMNIST \cite{41}.

\subsubsection{Preprocessing} The dataset was filtered to keep only the 82,698 images that contained specimen (removing e.g. nearly-all-black or -blue images, which are likely off the slide or pictures of marker on the slide). Regions surrounding tissue or corresponding to tears in the tissue were thresholded to white. Hue and lightness were obtained as the H and L channels from the standard HLS colorspace using the Python imagine library Pillow (v8.2.0).

Similarity $s_{ij}$ between each pair of images $i$ and $j$ was determined as the minimum of the similarities of four component features: hue, lightness, texture (e.g. sheets of nuclei), and structure (e.g. colonic crypts), as detailed below. This expert-feature engineering approach was used as it provided rapid proof of principle based on expert knowledge with a minimal set of human-expert examples (as opposed to a more involved/more data-hungry machine-learning approach).

\subsubsection{Hue component} The hue component was the overlap between hue histograms. 
Lightness component. The lightness component was the overlap between lightness histograms between 0.2 and 1.0, binned into 28 bins.

\subsubsection{Texture component} The texture component was a function of the similarity of mottledness (e.g. nuclei, sheets of cells) and stripedness (e.g. muscle fibers). The mottledness $m_i$ of an image $i$ was defined as the mean over all pixels of the maximum lightness difference $\delta_{\max⁡i}$ between a pixel and each of its eight nearest neighbors (normalized by the range of observed values, which happens to be 0.39): $\bar{m_i}=\delta_{\max}/0.39$ (where the bar indicates averaging). The mottledness similarity $m_{ij}$ between images $i$ and $j$ was then defined as $m_{ij}=0.1$ if $m_i<0.5$ and $m_j>0.5$, and $m_{ij}=1-|m_i-m_j|$ otherwise. Stripedness of an image $i$ was calculated in four directions: horizontal, vertical, and the two diagonals ($\sigma_{\uparrow i}$, $\sigma_{\rightarrow i}$, $\sigma_{\nwarrow i}$, $\sigma_{\nearrow i}$, respectively). In each case, it was defined as one minus the mean over all pixels of the sum of the differences $\delta_i$ between a pixel $i$ and its nearest neighbor in each direction (e.g., for horizontal stripedness, its neighbors to the left and right) times the sum $\delta_{\perp i}$ of the neighbors in the perpendicular direction (for horizontal stripedness, its neighbors above and below), again normalized by the range of observed values: $\sigma_{\_i}=(1-\bar{\delta}_i) \bar{\delta}/c_\perp$ for each direction (where \_ is replaced by $\uparrow, \rightarrow, \nwarrow, \text{or} \nearrow$ and the normalization constants were $c_\uparrow=0.229$, $c_\rightarrow=0.240$, and $c_\nwarrow=c_\nearrow=0.226$). The stripedness for the image as a whole, $\sigma_i$, was defined as the max of $\sigma_{\uparrow i}, \sigma_{\rightarrow i}, \sigma_{\nwarrow i}, \text{and} \sigma_{\nearrow i}$. The stripedness similarity $\sigma_{ij}$ was defined similarly to mottledness similarity: $\sigma_{ij}=0.1$ if $\sigma_{i}<0.5$ and $\sigma_{j}>0.5$, and $\sigma_{ij}=1-|\sigma_i-\sigma_j |$ otherwise. The texture similarity was then defined as $t_{ij}=\min⁡(m_{ij},\sigma_{ij})$.

\bigskip
\subsubsection{Structure component} A median filter was applied to each image to ignore outlier pixels. The structure component was then defined as the minimum of two subcomponents: similarity of structure size $S_{ij}$ and similarity of number of ``holes'' in the image $H_{ij}$.

$S_{ij}$ was determined as follows. If the resulting image had nearly uniform lightness (lightness range <0.5), $S_{ij}$ was ignored (because meaningful structures could not be easily extracted). For remaining images, structure size was determined by segmenting to find the largest patch $P_i$ around the darkest pixel in each image $i$. Thereafter, the following conditional logic was applied. If the largest structure in both images in the pair was <30 pixels, $S_{ij}$ was again ignored (structures were likely noise). Otherwise, the difference in lightness of the darkest pixel of each image was calculated ($\Delta$). If $\Delta$ was >0.2 times the lightness range of image $i$ and if $P_i$ was <30 pixels, $S_{ij}$ was again ignored; if $P_i$ was $\geq30$ pixels, $S_{ij}=0$. If instead $\Delta$ was $\leq0.2$ times the lightness range of image, and if in addition over half the pixels in each image had lightness >0.8 (e.g. indicating fat), $S_{ij}=0.9$. Otherwise, if over 60 percent of the pixels in one of the two images had lightness >0.8 (meaning one was very light and the other very dark), $S_{ij}=0.1$. Otherwise, $S_{ij}$ was set to be the size ratio of the smaller to the larger patch.

$H_{ij}$ was determined as follows. For each image $i$, the mask used to find $P_i$ was examined to count the number of light patches. (Patches that were only a single pixel in size were considered noise and ignored.) Each light patch was considered a hole. There was always at least 1 hole. $H_{ij}$ was defined as the ratio of the smaller to the larger number of holes for the pair of images. Structure similarity was defined as $\min⁡(S_{ij},H_{ij})$.

\bigskip
\subsubsection{Comparison to human experts} 100 image pairs were selected representing the range of similarities according to the above similarity function (from 0 to 1) to and presented independently to two practicing pathologists. These domain experts were instructed to assign the similarity between the two images, without further instruction, on a 1 to 10 scale. $R^2$ was calculated between their similarity scores as a measure of inter-expert agreement, and separately between each of their scores and $S_{ij}$.

\section{Results}

\subsection{Python package overview and main features}

The \textit{sentropy} package was developed to simplify the calculation of S-entropy (and traditional entropy) for Python users. It has the ability to calculate all frequency- and similarity-sensitive diversity measures presented in Reeve et al.’s work \cite{15} (S-entropies), as well as their similarity-insensitive counterparts (``traditional'' entropies), both in effective-number (preferred) and traditional forms. It offers command-line execution as a Python module for similarity-sensitive diversity calculations using a similarity matrix stored on disk and also provides a convenient interface when imported as a Python package.

\subsection{Availability and installation}

The \textit{sentropy} package is available via the Python Package Index (http://pypi.org) and on GitHub at https://github.com/ ArnaoutLab/sentropy. It can be installed by running

\begin{verbatim}
pip install sentropy     
\end{verbatim}

\noindent from the command line. The test suite runs successfully on Macintosh, Windows, and Unix systems. The unit tests and coverage report can also be run by 

\begin{verbatim}
pytest --pyargs sentropy --cov sentropy
\end{verbatim}

\subsection{Basic usage: alpha diversities}

We illustrate the basic usage of \textit{sentropy} on simple datasets of fruits (Fig. \ref{fig:1}) and animals (Fig. \ref{fig:2}) because these are discipline-agnostic and easy to interpret. We will use the terms ``subset'' or (``class'') and ``overall'' instead of ``subcommunity'' and ``metacommunity'' \cite{7}. (Note, when there is only a single dataset under study, the subset is the same as overall.) We format the input as pandas dataframes for clarity, but this is optional.

First, consider two datasets, each with 35 total elements. Each dataset has the same $n=6$ unique elements, each a type of fruit: apples, oranges, bananas, pears, blueberries, and grapes (Fig. \ref{fig:1}). Dataset 1a is mostly apples (Fig. \ref{fig:1}, left), while Dataset 1b has almost identical frequencies of all the fruits (Fig. \ref{fig:1}, right; Table \ref{table:2}).

\begin{table}[tb]
\begin{center}
\includegraphics[width=0.5\textwidth]{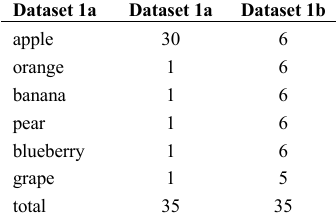}
\end{center}
\caption{
Datasets 1a\&1b: Fruits.
}
\label{table:2}
\end{table}

We wish to apply \textit{sentropy} to these datasets, being sensitive to the essential difference between them: the difference in frequencies of the unique elements. To do so, we first specify a species counts table, formatted as in Table \ref{table:2}. Then we pass it to the \verb|sentropy| function:

\begin{verbatim}
import pandas as pd
import numpy as np
from sentropy import sentropy

MEASURES =["alpha", "rho", "beta", "gamma", "normalized_alpha",\
"normalized_rho", "normalized_beta", "rho_hat", "beta_hat"]

counts_1a = {"dataset_1a": np.array([30, 1, 1, 1, 1, 1])}
counts_1b = {"dataset_1b": np.array([ 6, 6, 6, 6, 6, 5])}
   
df = sentropy(counts_1a, q=[0,1,np.inf], measure= MEASURES,\
level="overall", return_dataframe=True)
\end{verbatim}

Here we requested to get diversity indices for 3 different viewpoint parameters, and all families of measure that \textit{sentropy} supports. If we print out the output dataframe, we get Table (\ref{table:3}).

\begin{table*}[tb]
\begin{adjustwidth}{-3cm}{}
\includegraphics[width=1.5\textwidth]{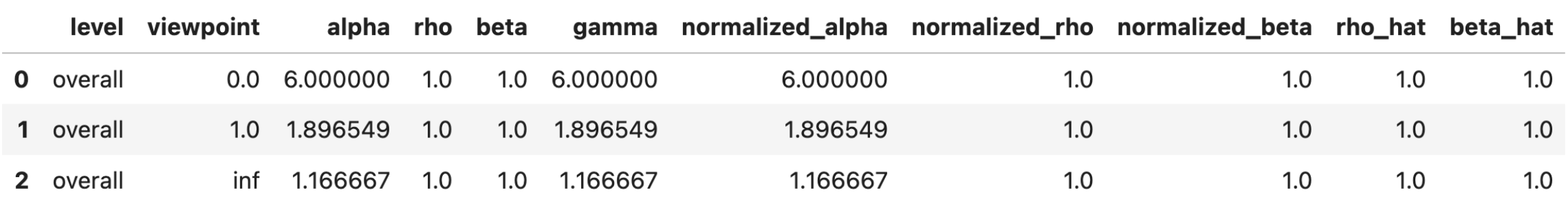}
\end{adjustwidth}
\caption{
\textit{sentropy} output for Dataset 1a.
}
\label{table:3}
\end{table*}
Looking at the alpha column, we see that the value of $D_{1}$ for this dataset is $1.90$. If we do not pass anything to \verb|q|, then only viewpoint $1$ is computed, and if we do not pass anything to \verb|measure|, then only $alpha$ diversity is computed. Instead of outputting a dataframe, it is also possible to have \verb|sentropy| output an object that we can query:
\begin{verbatim}
D = sentropy(counts_1a, q=[0,1,np.inf], level="both")
D(which='overall', q=0, measure='alpha')
\end{verbatim}
and we get the output $6$, because there are 6 species of fruits. We can optionally omit the \verb|measure| argument in the above, because the sentropy object only computed alpha diversity anyway. By default, \verb|level| takes value \verb|'overall'|, which means only the set-level diversities will be computed. When we query the sentropy object, we can pass to the \verb|which| argument the name of the subset of interest, or \verb|which="overall"| if we are interested in diversity at that level. Also, if we had passed a numpy array instead of a pandas dataframe as the abundance matrix, then the names of the subsets are simply their indices. In that case we can pass an integer representing the index of the subset to \verb|which|.

If furthermore we are interested in the traditional forms of the LCR diversity indices---logarithms (with base \textit{e})---instead of effective numbers, we can pass \verb|eff_no=False|. By default, the \verb|eff_no| argument (which stands for effective number) takes value True. So, if we run:
\begin{verbatim}
H = sentropy(counts_1a, q=0, eff_no=False)
print(H)
\end{verbatim}
we get the answer $1.79$. Note that in this case where we want to compute only one viewpoint parameter, one measure and there is one subset in the set, \verb|sentropy| returns a number rather than an object, because there is only one diversity index to be computed.

If we now repeat the computation for Dataset 1b, we find that $D_1\approx5.99$ for that dataset. The larger value of $D_1$ for Dataset 1b aligns with the intuitive sense that more balance in the frequencies of unique elements means a more diverse dataset.

\textit{sentropy} can also calculate S-entropy measures for any user-supplied definition of similarity. To illustrate, we now consider a second example in which the elements of two datasets are all unique: here, animals (Fig. \ref{fig:2}). Uniqueness means the frequency distributions of the two datasets are identical, so similarity is the only factor that can influence our sense of whether one or the other dataset is more diverse. The investigator always has a choice of similarity measure, which can be choosen to investigate the question at hand; here we consider (approximate) phylogenetic similarity. Dataset 2a consists entirely of birds, so all entries in the similarity matrix are close to 1:

\begin{verbatim}
labels_2a = [
    "owl", "eagle", "flamingo", "swan", 
    "duck", "chicken", "turkey", "dodo", 
    "dove"
    ]
no_species_2a = len(labels_2a)
S_2a = np.identity(n=no_species_2a)
S_2a[0][1:9] = (0.91, 0.88, 0.88, 0.88, 0.88, 0.88, 0.88, 0.88) # owl
S_2a[1][2:9] = (0.88, 0.89, 0.88, 0.88, 0.88, 0.89, 0.88) # eagle
S_2a[2][3:9] = (0.90, 0.89, 0.88, 0.88, 0.88, 0.89) # flamingo
S_2a[3][4:9] = (0.92, 0.90, 0.89, 0.88, 0.88) # swan
S_2a[4][5:9] = (0.91, 0.89, 0.88, 0.88) # duck
S_2a[5][6:9] = (0.92, 0.88, 0.88) # chicken
S_2a[6][7:9] = (0.89, 0.88) # turkey
S_2a[7][8:9] = (0.88) # dodo
# dove
S_2a = np.maximum( S_2a, S_2a.transpose() )
S_2a = pd.DataFrame({labels_2a[i]: S_2a[i] for i in range(no_species_2a)}, 
                    index=labels_2a)
\end{verbatim}

We make a DataFrame of counts in the same way as in the previous example:

\begin{verbatim}
counts_2a = pd.DataFrame(
    {"dataset_2a": [1, 1, 1, 1, 1, 1, 1, 1, 1]}, index=labels_2a)
\end{verbatim}

To compute similarity-sensitive diversity measures, we now pass the similarity matrix to the \verb|similarity| argument of \verb|sentropy|:

\begin{verbatim}
result_2a = sentropy(counts_2a, similarity=S_2a, q=0)
print(result_2a)
\end{verbatim}

The output tells us that $D_1^Z\approx1.11$. The fact that this number is close to 1 reflects the fact that all individuals in this dataset are phylogenetically very similar to each other: they are all birds. In contrast, Dataset 2b (Fig. \ref{fig:2}) consists of members from two different phyla: vertebrates and invertebrates. As above, we define a similarity matrix:

\begin{verbatim}
labels_2b = ("ladybug", "bee", "butterfly", "lobster", "fish", 
             "turtle", "parrot", "llama", "orangutan")
no_species_2b = len(labels_2b)
S_2b = np.identity(n=no_species_2b)
S_2b[0][1:9] = (0.60, 0.55, 0.45, 0.25, 0.22, 0.23, 0.18, 0.16) # ladybug
S_2b[1][2:9] = (0.60, 0.48, 0.22, 0.23, 0.21, 0.16, 0.14) # bee
S_2b[2][3:9] = (0.42, 0.27, 0.20, 0.22, 0.17, 0.15) # butterfly
S_2b[3][4:9] = (0.28, 0.26, 0.26, 0.20, 0.18) # lobster
S_2b[4][5:9] = (0.75, 0.70, 0.66, 0.63) # fish
S_2b[5][6:9] = (0.85, 0.70, 0.70) # turtle
S_2b[6][7:9] = (0.75, 0.72) # parrot
S_2b[7][8:9] = (0.85) # llama
#orangutan
S_2b = np.maximum( S_2b, S_2b.transpose() )
S_2b = pd.DataFrame({labels_2b[i]: S_2b[i] for i in range(no_species_2b)}, 
                    index=labels_2b)
\end{verbatim}

The values of the similarity matrix indicate high similarity among the vertebrates, high similarity among the invertebrates, and low similarity between vertebrates and invertebrates. To calculate the alpha diversity (with $q=1$ as above), we proceed as before:

\begin{verbatim}
counts_2b = pd.DataFrame({"dataset_2b": [1, 1, 1, 1, 1, 1, 1, 1, 1]}, 
                        index=labels_2b)
result_2b = sentropy(counts_2b, similarity=S_2b, q=0)
print(result_2b)
\end{verbatim}

Inspecting the result, we find $D_1^Z\approx2.16$. That this number is close to 2 reflects the fact that members in this dataset belong to two broad classes of animals: vertebrates and invertebrates. The small difference compared to 2 is interpreted as the contribution of the diversity within each phylum. Note that if we had instead simply placed animals into two bins, vertebrates and invertebrates, this contribution would be lost.

\subsection{Basic usage: beta diversities}

Recall that beta diversity is between-group diversity. We can use \textit{sentropy} to compare the diversity of the vertebrates and invertebrates. To do so, we define two classes within Dataset 2b---invertebrates and vertebrates---as follows:

\begin{verbatim}
counts_2b_1 = pd.DataFrame(
    {"invertebrates": [1, 1, 1, 1, 0, 0, 0, 0, 0], # invertebrates
    "vertebrates": [0, 0, 0, 0, 1, 1, 1, 1, 1], #   vertebrates
    }, index=labels_2b)
\end{verbatim}

We then obtain the representativeness  $\bar{\rho}$ of each subset, here at $q=0$, as follows:
\begin{verbatim}
result_2b_1 = sentropy(counts_2b_1, similarity=S_2b, q=0, \
measure=['alpha', 'normalized_rho'], level="class")
result_2b_1(which = "invertebrates", measure='normalized_rho')
\end{verbatim}
and 
\begin{verbatim}
result_2b_1(which = "vertebrates", measure='normalized_rho')
\end{verbatim}
We find the answers $0.63$ for the invertebrates and $0.67$ for the vertebrates. Recall $\bar{\rho}$ indicates how well a class represents the overall dataset. Note we can also ask which subset is more diverse by calculating the alpha diversities of the two classes (also at $q=0$, for ease of comparison):
\begin{verbatim}
result_2b_1(which = "invertebrates", measure='alpha')
\end{verbatim}
and 
\begin{verbatim}
result_2b_1(which = "vertebrates", measure='alpha')
\end{verbatim}
We find the answers $3.54$ for the invertebrates and $2.30$ for the vertebrates. Thus, the invertebrates are more diverse than the vertebrates. In contrast, suppose we split Dataset 2b into two subsets at random, without regard to phylum:
\begin{verbatim}
counts_2b_2 = pd.DataFrame(
{
   "random_subset_1": [1, 0, 1, 0, 1, 0, 1, 0, 1],
   "random_subset_2": [0, 1, 0, 1, 0, 1, 0, 1, 0],
},
index=labels_2b
)
\end{verbatim}
Proceeding again as above, we query the sentropy object to obtain the representativeness of the 2 subsets, with the command
\begin{verbatim}
result_2b_2 = sentropy(counts_2b_2, similarity=S_2b, q=0, \
measure='normalized_rho', level="class")
result_2b_2(which="random_subset_1")
\end{verbatim}
and the command
\begin{verbatim}
result_2b_2(which="random_subset_2")
\end{verbatim}
We find answers $0.93$  and $0.92$ respectively. These values are higher than the corresponding ones using abundance matrix 2b\_1, which reflects the fact that each group now has a roughly equal mix of vertebrates and invertebrates.

\subsection{Advanced usage: similarity matrix format}

The similarity matrix format---DataFrame, memmap, filepath, or function---should be chosen based on the use case, with particular attention to dataset size. Our recommendations: 

\begin{itemize}
    \item If the similarity matrix fits in RAM, pass it as a pandas.DataFrame or numpy.ndarray
    \item 	If the similarity matrix does not fit in RAM but does fit on your hard drive (HD), pass it as a cvs/tsv filepath or numpy.memmap. To illustrate passing a csv file, we re-use \verb|counts_2b_1| and \verb|S_2b| from above and save the latter as .csv files (note index=False, since the csv files do not contain row labels).

\begin{verbatim}
S_2b.to_csv("S_2b.csv", index=False)
\end{verbatim}

Then we can pass the path as a string to \verb|sentropy|:

\begin{verbatim}
sentropy(counts_2b_1, similarity='S_2b.csv', chunk_size=5, \
q=0)
\end{verbatim}

We can optionally use the \verb|chunk_size| argument to specify how many rows of the similarity matrix are read from the file at a time.

    \item 	If the similarity matrix does not fit in either RAM or HD, pass a similarity function to the \verb|similarity| argument and the feature set that will be used to calculate similarities to the \verb|sfargs| argument of \verb|sentropy|. Here's an example with a set of 2 amino acid sequences:

    \begin{verbatim}
    from polyleven import levenshtein as edit_distance

    def similarity_function(species_i, species_j):
        return 0.3**edit_distance(species_i, species_j)
    
    test_seqs = ['CARDYW', 'CARDYV']
    test_nos = [10, 1]
    
    test = pd.DataFrame(
        {"test_nos": test_nos},
        index=test_seqs
        )
    
    sentropy(
        test, 
        similarity=similarity_function,
        sfargs=np.array(test_seqs),
        q=0
        )
    \end{verbatim}
    
    and we get the answer $1.22$ for the alpha diversity. \verb|sentropy| admits the optional argument \verb|chunk_size|, which determines how many rows of the similarity matrix are processed at a time. Note that construction of the similarity matrix is an $\mathcal{O}(N^2)$ operation; if your similarity function is expensive, this calculation can take time for large datasets. If the similarity callable is a symmetric function (as most similarities are), then a twofold speedup can be obtained by passing \verb|symmetric=True| to \verb|sentropy|.
    
    If we additionally pass \verb|parallelize=True|, then the computation of diversity indices will be parallelized using the Ray package (all available processing cores will be utilized). In this case we can also pass a number to the argument \verb|max_inflight_tasks| to specify at most how many parallel tasks should be submitted to Ray at a time (to avoid overwhelming Ray).

\end{itemize}

\subsection{Advanced usage: inter-set ordinariness}

The ordinariness of species $i$ is the total abundance of all the species in the dataset that are similar to $i$, weighted by their similarity and their abundance. (This includes $i$ itself, which is of course 100\% similar to itself; $Z_{ii} = 1$.) If the dataset contains many species that are similar to $i$ and/or if those species are very abundant, the ordinariness of species $i$ will be high: species $i$ will be considered unexceptional relative to these many/abundant similar species in the dataset, and in that sense is quite "ordinary" for the dataset.

There are situations in which one is interested in the ordinariness of a species $j$ that is \textit{not} in the dataset: i.e. how similar the species in the subset are to this outside member, weighted by their relative abundances and their similarity to $j$. (See \cite{braun2023contrasting} for an example, in which the investigators start with an antibody $j$ and wish to calculate its ordinariness in antibody repertoires. In this context, the ordinariness is the repertoire's ``binding capacity'' for the molecules that antibody $j$ is best at binding.)

\verb|sentropy| can calculate ordinariness, whether the species of interest is present in the subset (i above) or not (j):

\begin{verbatim}
from sentropy.abundance import Abundance
from sentropy.ray import IntersetSimilarityFromRayFunction

def similarity_function(species_i, species_j):
    return 1 / (1 + np.linalg.norm(species_i - species_j))

counts = np.array([[1,0],[0,1],[1,1]])
abundance = Abundance(counts, subsets_names=['1', '2'])
community_species = np.array([[1, 2], [3, 4], [5, 6]])
query_species = np.array([[1, -1], [3, 6]])
similarity = IntersetSimilarityFromRayFunction(
  similarity_function,
  query_species,
  community_species)
ordinariness = similarity @ abundance
\end{verbatim}

The code above computes the ordinariness of each of the 2 query species with respect to a dataset made of 3 other species. Here the similarity function is evaluated on the fly, with Ray parallelization.

\subsection{Advanced usage: Kullback-Leibler divergence and its cousins}

\verb|sentropy| can also compute the Kullback-Leibler divergence (i.e. relative entropy) between datasets, as well as its generalizations the Rényi divergences and their S-entropic cousins. To do so, we simply pass 2 abundance arrays as the first 2 arguments of \verb|sentropy|:
\begin{verbatim}
sentropy(counts_2b_1, counts_2b_2, similarity=S_2b, q=1, \
return_dataframe=True, level="both")
\end{verbatim}
we get a tuple of 2 elements, the first of which is a float representing the superset Rényi divergence at viewpoint 1 (in this case, 1), and the second of which is a DataFrame containing Rényi divergences between pairs of subsets, in this case:
\begin{table}[htbp]
\centering
\begin{tabular}{|l|c|c|}
\hline
 & subset\_2b\_3 & subset\_2b\_4 \\
\hline
subset\_2b\_1 & 1.66 & 1.55 \\
subset\_2b\_2 & 1.43 & 1.56 \\
\hline
\end{tabular}
\caption{Relative S-entropies between subsets of Dataset 2b\_1 and Dataset 2b\_2.}
\label{tab:KLdiv}
\end{table}

\subsection{Advanced usage: PyTorch and GPU support}
For heavier computations, it is possible to use PyTorch instead of numpy to obtain some acceleration. To do so, we pass \verb|backend="torch"| to \verb|sentropy|. (By the fault, the backend argument takes value \verb|"numpy"|.) In order to have the computation of the diversity indices run on the GPU, we can additionally pass \verb|device="mps"| or \verb|device="cuda"|, depending on whether the computation runs on a Mac computer with Apple silicon, or a computer with NVIDIA CUDA. (By default, the device argument takes value \verb|cpu|, which means the computation runs on the CPU.) For example, consider the following set with 100 subsets and 10000 entities:

\begin{verbatim}
big_counts = np.random.randint(101, size=(10000,100))

n = 10000
indices = np.triu_indices(n, k=1)
values = np.random.rand(len(indices[0]))
big_sim_matrix = np.zeros((n, n))
big_sim_matrix[indices] = values
big_sim_matrix = big_sim_matrix + big_sim_matrix.T
np.fill_diagonal(big_sim_matrix, 1.0)
\end{verbatim}

On a Mac computer with MPS, we can call:
\begin{verbatim}
sentropy(big_counts, similarity=big_sim_matrix, \
q=[0,1, 1.5, np.inf], measure=MEASURES, backend='torch', device='mps')
\end{verbatim}
which we found to be around 3 times faster than the CPU-bound computation without torch. If we only pass \verb|backend='torch'| without passing \verb|device='mps'|, then the computation will utilize torch but will run on the CPU.

\subsection{Command-line usage}

The \textit{sentropy} package can also be used from the command line as a module (via \verb|python -m|). To illustrate this use of \textit{sentropy}, we re-use again the example with \verb|counts_2b_1| and \verb|S_2b|, now with \verb|counts_2b_1| also saved as .csv files (note again \verb|index=False|):

\begin{verbatim}
counts_2b_1.to_csv("counts_2b_1.csv", index=False)
\end{verbatim}

Then from the command line: 

\begin{verbatim}
python -m sentropy -i counts_2b_1.csv -s S_2b.csv -qs 0 1 inf
\end{verbatim}

The output is a table with all the diversity measures for $q=0$, 1, and $\infty$. Note that while .csv or .tsv are acceptable as input, the output is always tab-delimited. The input filepath (\verb|-i|) and the similarity matrix filepath (\verb|-s|) can be URLs to data files hosted on the web. Also note that values of $q>100$ are all calculated as $q=\infty$.

To compute relative entropies in the terminal, we simply pass a second csv file for the other abundance matrix:

\begin{verbatim}
python -m sentropy -i counts_2b_1.csv counts_2b_2.csv -s S_2b.csv -qs 1
\end{verbatim}

The full list of flags is -i (for input filepath), -o (for output filepath), -s (for the filepath to the similarity matrix), -qs (for the viewpoint parameters), -ms (for the diversity measures of interest), -chunk\_size (for the chunk size when reading the file into memory), -level (for whether to compute diversities at the overall/subset level, or both), -eff\_no (for whether to compute effective numbers or entropies), -backend (for whether to use numpy or torch), -device (for whether to use the CPU or the GPU). For further options, consult the help:

\begin{verbatim}
python -m sentropy -h
\end{verbatim}

\subsection{Applications}

We now illustrate some practical applications of \textit{sentropy} using sample datasets from the fields of immunomics, metagenomics, medical imaging, and computational pathology, as examples of the many types of dataset to which this package can be usefully applied. Jupyter notebooks with these examples are available via the \textit{sentropy} GitHub repository.

\bigskip
\subsubsection{Immunomics}

\begin{figure*}[t]
\centering
  \includegraphics[width=1.\textwidth]{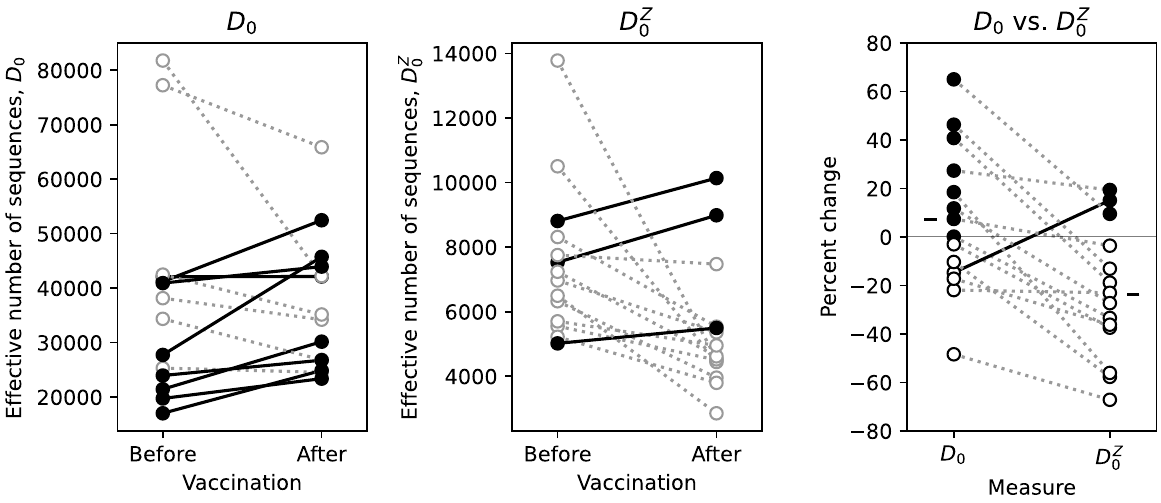}
  \caption{
  Immunomics: Similarity-insensitive vs. -sensitive measures in influenza vaccination. Diversity of IGH CDR3 immunomes according to the similarity-insensitive measure $D_0$ (left) and its similarity-sensitive counterpart $D_0^Z$ (middle), and a comparison of the two (right) before vs. after influenza vaccination. Left and middle: light/dotted lines denote subjects where diversity falls. Right: each line/pair of symbols show $D_0$ and $D_0^Z$ for the same subject. Dark line shows the one subject where vaccination was associated with fewer, more different sequences. Dashes in the margins indicate averages of $D_0$ and $D_0^Z$.
  }
  \label{fig:3}
\end{figure*}

Antibody (IG) and T-cell receptor (TR) repertoires are famously diverse, with tens of thousands of different IG and TR gene sequences in every milliliter of human blood \cite{16}, \cite{20}, \cite{23}, \cite{24}, \cite{25}. The frequency of a given sequence rises with exposure to antigens as cells divide, forming clones of identical or similar sequences, and falls to baseline over time as cells die. As a result, measures that are both frequency-sensitive and similarity-sensitive are useful for describing the state of the adaptive immune system \cite{10}, \cite{16}, \cite{26}.

To illustrate, we revisit a dataset of 30 IG heavy-chain (IGH) repertoires from 15 subjects taken before and after influenza vaccination \cite{27}, each subsampled to a same-sized 100,000-sequence subset \cite{10}. Vaccination is associated with an increase in the number of sequences---$D_0$, which is similarity insensitive---interpreted as vaccination leading to new sequences \cite{27}, a finding confirmed in a recent re-analysis of the data \cite{10} that corrected for sampling bias (using the recon package \cite{16}) (Fig. \ref{fig:3}a). We can use sentropy to measure the alpha diversity for each before-and-after pair using the S-entropy measure $D_0^Z$, using binding similarity as previously described to construct the similarity matrix \cite{10}. Unlike $D_0$, which generally rises as antigen-responsive cells divide and accumulate mutations, $D_0^Z$ falls in 12 of the 15 subjects (Fig. \ref{fig:3}b). The combination of these two trends tells us that new sequences that are seen following vaccination are likely related, consistent with expansion of pre-existing clones. The three exceptions suggest something different is happening immunologically in these subjects, perhaps new immune responses or fewer mutations.

\begin{figure}[t]
  \centering
  \includegraphics[width=0.4\textwidth]{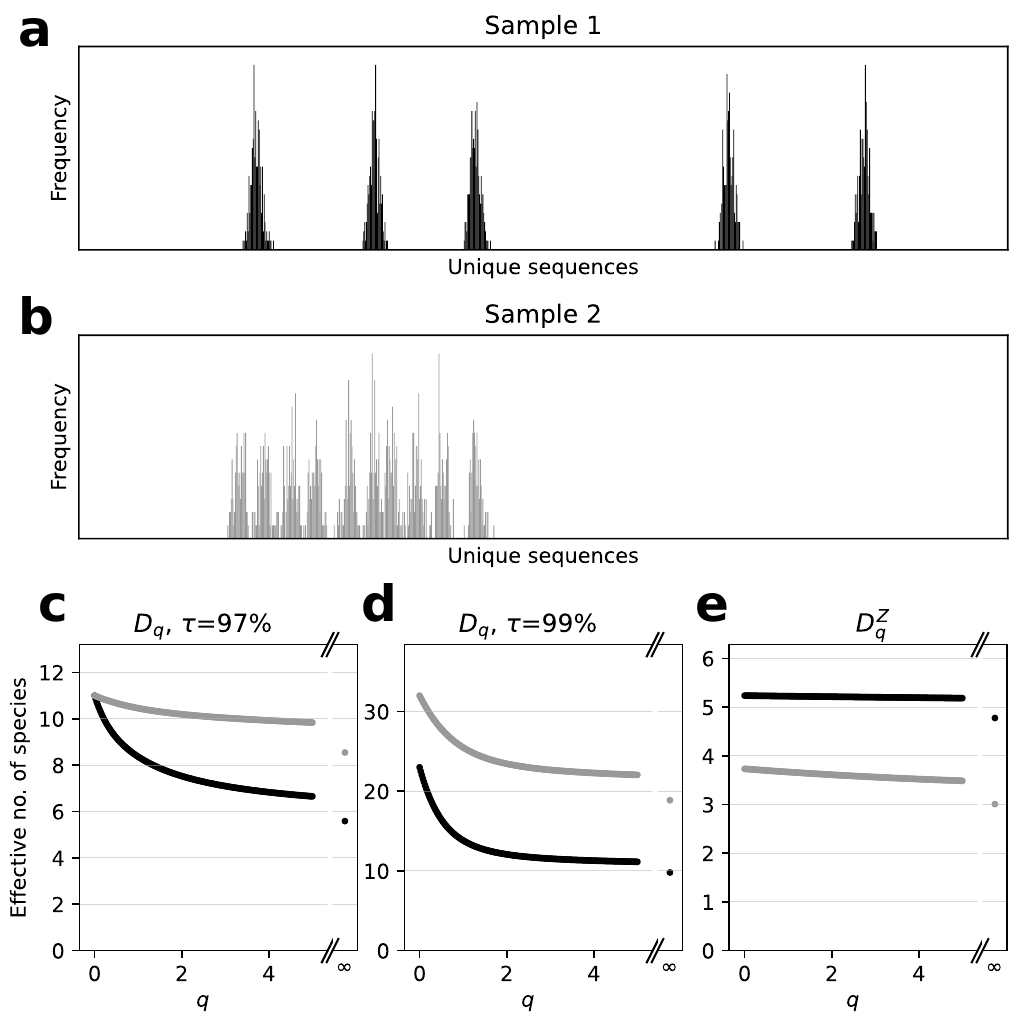}
  \caption{
  Metagenomics: Traditional vs. S-entropy alpha measures in synthetic samples. Sample 1 (a) and 2 (b) both have 1,000 sequences. For ease of illustration, distance along the x-axis is proportional to similarity (i.e. more similar sequences are nearer each other). In Sample 1, sequences form five highly distinct clusters. In Sample 2, sequences form 10 more similar clusters. The distribution of sequences within each cluster is Gaussian, consistent with observations from the Human Microbiome Project [11]. Using binning to account for similarity and then measuring diversity using traditional/similarity-insensitive measures ($D_q$), diversity depends on the binning threshold $\tau$ and frequency weighting q. (c) At $\tau=97\%$ and $q=0$, the two samples are equally diverse, with $D_0=11$ species; At higher $q$, Sample 2 is more diverse, with diversities falling to $D_\infty=8.6$ vs. $D_\infty=5.6$ species, respectively, at $q=\infty$). (d) At $\tau=99\%$, both samples are much more diverse, with $D_0=32$ vs. 23 species, respectively. (Note that at this $\tau$, Sample 2 is more diverse than Sample 1 for all q.) (e) In contrast, accounting for similarity using S-entropy measures ($D_q^Z$), which avoids the need for binning, the order flips: Sample 1 is now more diverse than Sample 2 (for all q), with $\sim5$ vs. $\sim3-4$ species, respectively, reflecting both the number and the grouping of the clusters. (Here similarity $s_{ij}$ between sequences $i$ and $j$ is calculated as $s_{ij}=e^{-k\Delta_{ij}}$, where $\Delta_{ij}$ is the Levenshtein distance between sequences $i$ and $j$ and $k=0.02$.)
  }
  \label{fig:4}
\end{figure}

\bigskip
\subsubsection{Metagenomics}

Measuring diversity is also of interest in metagenomics, where microbiomes from e.g. stool or soil samples are high-throughput sequenced to determine the number and frequency of species present. For practical reasons, often the sequenced material is not a complete genomes but telltale genes, such as 16S rRNA sequences, or fragments of such genes \cite{28}.

This task is challenging for three reasons. First, the concept of ``species'' is not as well defined for bacteria as it is for more complex organisms. Second, the similarity of a pair of species, measured in terms of phylogeny or sequence identity, can differ substantially from pair to pair, such that a pair of organisms with a given degree of sequence identity might be classified as the same species in one genus but as distinct species---or even distinct genera---in another. Third, many recovered sequences are difficult to map uniquely to known species. (Tools such as Unifrac that characterize the similarity between microbiomes according to the proportion of shared lineages still require confidence in lineage assignment \cite{29}.) These three challenges are in addition to the general challenge faced in measuring the diversity of complex populations: assuring that the diversity of the sample is an accurate representation of the diversity of the population from which the sample was drawn \cite{16}.

To address these three challenges, a common first step in measuring microbiome diversity has been to bin sequences into OTUs based on a set threshold level of sequence similarity, e.g. 95\% or 97\%. Unfortunately, binning is not without its own challenges \cite{11}, \cite{12}. First is the problem of ``adverse triplets'' \cite{12}: sequences $i$, $j$, and $k$ for which the similarity between $i$ and $j$ is above the threshold, the similarity between $j$ and $k$ is also above the threshold, but the similarity between $i$ and $k$ is below the threshold. Binning all three into the same OTU because $i$-$j$ and $j$-$k$ are above the threshold results in the OTU containing $i$ and $k$ even though their similarity is below the threshold. A second challenge is determining what the appropriate threshold should be \cite{11}, \cite{12}. These are general problems with binning or other forms of unique assignment. I.e. they are not specific to metagenomics: they are a necessary consequence of discretization of continuous or near-continuous variables, such as sequence similarity, into discrete bins.

\begin{figure*}[t]
    \centering
    \includegraphics[width=1\textwidth]{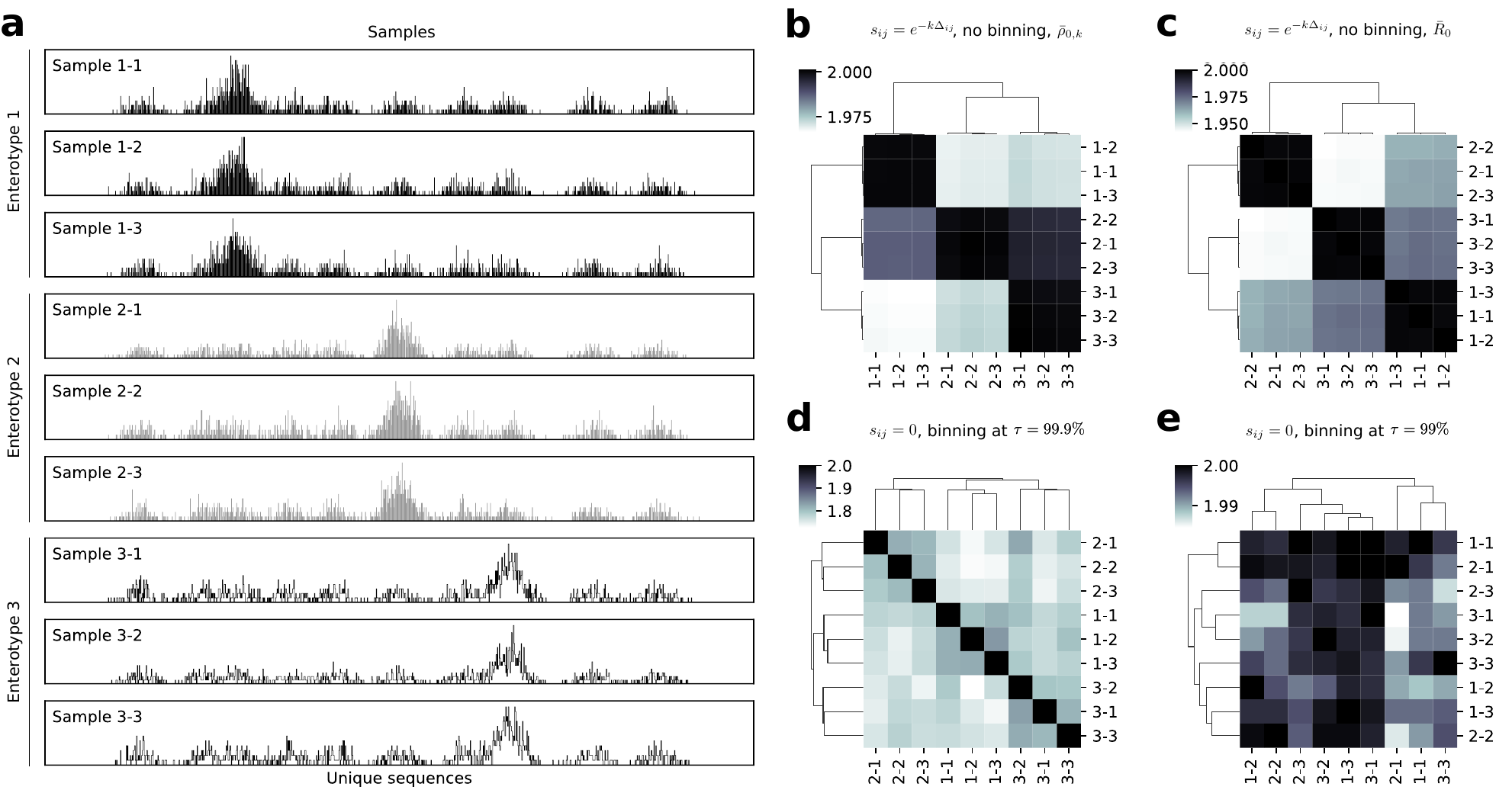}
    \caption{
    Metagenomics: Beta diversities to elucidate population structure in synthetic samples. (a) Nine synthetic samples created as in Fig. 4a-b, three representing each of three enterotypes: black, Enterotype 1 (Samples 1-1 to 1-3); gray, Enterotype 2 (Samples 2-1 to 2-3); white, Enterotype 3 (Samples 3-1 to 3-3). (b)-(c) Clustering using similarity $s_{ij}$ between sequences i and j as defined in Fig. \ref{fig:4} without binning, according to (b) the $q=0$ representativeness ($\bar{\rho}_0$) of the $k^\text{th}$ sample for the pair of samples indicated by the heatmap cell and (c) the average representativeness of each member of the pair ($\bar{R}_0$).
    }
    \label{fig:5}
    \end{figure*}

S-entropy offer an alternative way to measure the diversity of microbiomes. We used \textit{sentropy} to illustrate using synthetic/in silico microbiome samples reminiscent of enterotypes in human gut microbiomes \cite{30} \cite{31}. An enterotype is a microbiome that exhibits a certain compositional pattern, e.g. enrichment in bacteria of the genus \textit{Bacteroides} \cite{30}, \cite{32}, \cite{33}, \cite{34}. Consider the two samples in Fig. \ref{fig:4}. Both have clusters of similar sequences; Sample 1 has five highly distinctive clusters (Fig. 4a) while Sample 2 has 10 clusters that are all fairly similar to each other (Fig. \ref{fig:4}b). Similarity can be accounted for in two ways: by binning and then using traditional entropy or by not binning and instead using S-entropy measures.

These yield different results. Binning sequences yields diversity values that depend on the choice of binning threshold but generally show Sample 2 as having higher diversity, reflecting the larger number of clusters (Fig. \ref{fig:4}c, d). In contrast, S-entropy shows Sample 1 as having higher diversity, reflecting the larger range that the sequences in Sample 2 span (Fig. \ref{fig:4}e). Specifically, $D_0^Z$ is a little over 5.4 in Sample 1, reflecting the five distinct clusters plus their (limited) intra-cluster diversity, whereas it is 3.6 in Sample 2, reflecting that the 10 clusters are themselves highly related (``effectively'' having the same diversity as a sample with 3.6 completely distinct sequences).

The \textit{sentropy} package is also useful for investigating population structure across different samples (subcommunities), for example to identify correlates of disease [35]. One way to do so is by defining the similarity $s_{ij}$ between each pair of unique sequences $i$ and $j$, treating each pair of samples as its own dataset, and then using $s_{ij}$ to calculate how well each sample $k$ represents the pair. This can be done either using the quantity, $\bar{\rho}_k$, which is the representativeness of sample $k$, or $\bar{R}$, which is the average representativeness over both samples in the pair. $\bar{\rho}_k$ generally differs for each sample in the pair, reflecting asymmetries in the overlap between them; in contrast $\bar{R}$ is always symmetric, and so should be thought of as an attribute of the pair, not of the constituent samples. Note $\bar{\rho}_k$ can be calculated at any $q$; in this example we omit the $q$ subscript for readability and use $q=0$ unless otherwise specified. \textit{sentropy} readily calculates $\bar{\rho}_k$ and $\bar{R}$ for any $q$. Once calculated, one can use these values to investigate population structure by clustering on the resulting matrix (Fig. \ref{fig:5}).

\begin{figure}[t]
\centering
\includegraphics[width=1.\textwidth]{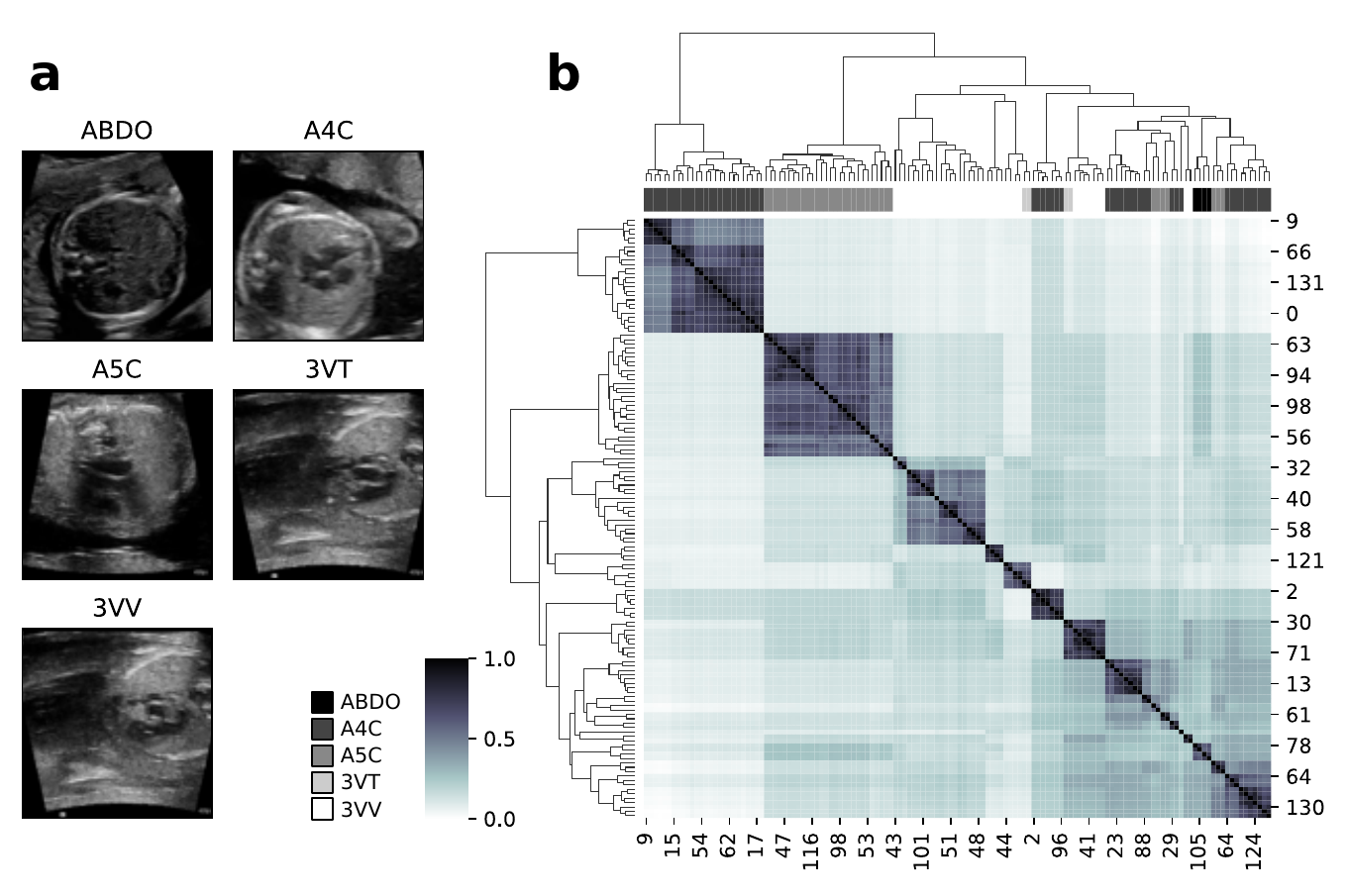}
\caption{
Medical imaging: Dataset structure. Example images from each labeled class (a) and heatmap of the similarity matrix for all 136 images (b) with several clusters of similar images clearly visible along the diagonal, often but not always correlating with the class label. The bar above the heatmap indicates the view (class label) of each image (ABDO, A4C, A5C, 3VT, 3VV). Similarity here is based on root-mean-square error (RMSE) of pixel differences between images i and j according to $s_{ij}=e^{-\text{RMSE}_{ij}}$. Identical images have a similarity of 1 and completely distinct images have a similarity approaching 0. Note that while RMSE is not invariant to rotation or translation, ultrasound images (like many medical images) have a privileged orientation and position.
}
\label{fig:6}
\end{figure}

For example, consider the nine synthetic samples in Fig. \ref{fig:5}a. Using the same $s_{ij}$ as in the previous example, clustering by either $\bar{rho}_k$ (Fig. \ref{fig:5}b) or $\bar{R}$ (Fig. \ref{fig:5}c) clearly shows these samples cluster into three enterotypes. Alternatively, one could instead use the OTU approach: account for similarity by choosing a sequence-similarity threshold $\tau$, bin according to that threshold, and then use \textit{sentropy} without $s_{ij}$ to calculate and cluster on $\bar{\rho}_k$. However, OTU results are highly sensitive to the choice of threshold (Fig. \ref{fig:5}d-e). The block structure is discernible only in the dendrogram in Fig. \ref{fig:5}d, whereas there is no obvious clustering apparent in Fig. \ref{fig:5}e. Note that both approaches require a choice about how similar two sequences are: in the former case via $s_{ij}$ and in the latter via $\tau$. (Recall that choosing $s_{ij}=1$ for sequences whose similarity is $\leq\tau$ and $s_{ij}=0$ otherwise makes the latter a special case of the former, except without the problem of adverse triplets.) The choice of $s_{ij}$ will depend on the context (sequence similarity, phylogenetic similarity, functional similarity, etc.).

\bigskip
\subsubsection{Medical Imaging}

Medical imaging is an active area of interest in machine learning (ML). Examples include view- and disease-classification tasks on X-rays, magnetic-resonance imaging (MRI) and positron-emission tomography (PET) scans, and ultrasounds (the most widely used and cost-effective modality for internal imaging). Training ML models on these and other types of medical images usually requires large, class-balanced medical-imaging datasets \cite{9}. This section illustrates use of the sentropy package to characterize such datasets, using an echocardiogram dataset as an example \cite{36}.

The dataset consists of 136 cardiac ultrasound images (echocardiograms) labeled for five classes according to which of five views of the heart they show (A4C, A5C, 3VV, ABDO, and 3VT), selected for ease of illustration (Fig. \ref{fig:6}). Unlike the previously discussed examples, in medical-imaging datasets each element generally appears only once, so all the images occur with the same frequency ($1/n$, where $n$ is the total number of images in the dataset, i.e. the dataset size). As a result, image frequency is generally not very informative in characterizing the dataset. However, similarity can be, as we show.

We suppose that only half the images in this dataset are already labeled. We want a sense of what might be gained through the effort of labeling the other half. We know the size of our labeled dataset would double. Assuming that ML benefits from datasets that are not only large but diverse, how much actual diversity would we add? We can answer this question by asking how well each half reflects the composition of the full dataset. The preceding section illustrated the utility of $\bar{\rho}_k$ for this purpose. In this section we will use a related measure introduced in the sentropy package: $\hat{\rho}_k$ (``rho-hat''), defined as $\hat{\rho}_k=(\rho_{k-1})/(N-1)$, where $N$ is the number of subsets. $\hat{\rho}_k$ has the convenient property of ranging from 0 to 1 (except for extreme/pathological cases), which may be more interpretable than the alternatives. A subset $k$ that is maximally distinctive relative to the full dataset will have $\hat{\rho}_k=0$ while a subset that is maximally redundant will have $\hat{\rho}_k=1$. The dataset-level analog of $\hat{\rho}_k$ is defined by simply by taking the power mean of the class values of $\hat{\rho}_k$, just like for any other diversity indices.

Similarly, we define $\hat{\beta}_k=(\beta_k N-1)/(N-1)$, which typically ranges from 0 to 1 (except in extreme/pathological cases). The overall-dataset analog of $\hat{\beta}_k$ is again defined to be the power mean of the class values. In pathological cases where some of the class values are negative, the overall $\hat{\beta}$ is ill-defined, since the power mean is only defined for positive values. Finally, we note the special case $N=1$. In this case, $\rho$ and $\beta$ are both 1, leading to division by zero, so we define $\hat{\rho}=\hat{\beta}=1$.

\begin{figure*}[t]
\centering
\includegraphics[width=1.\textwidth]{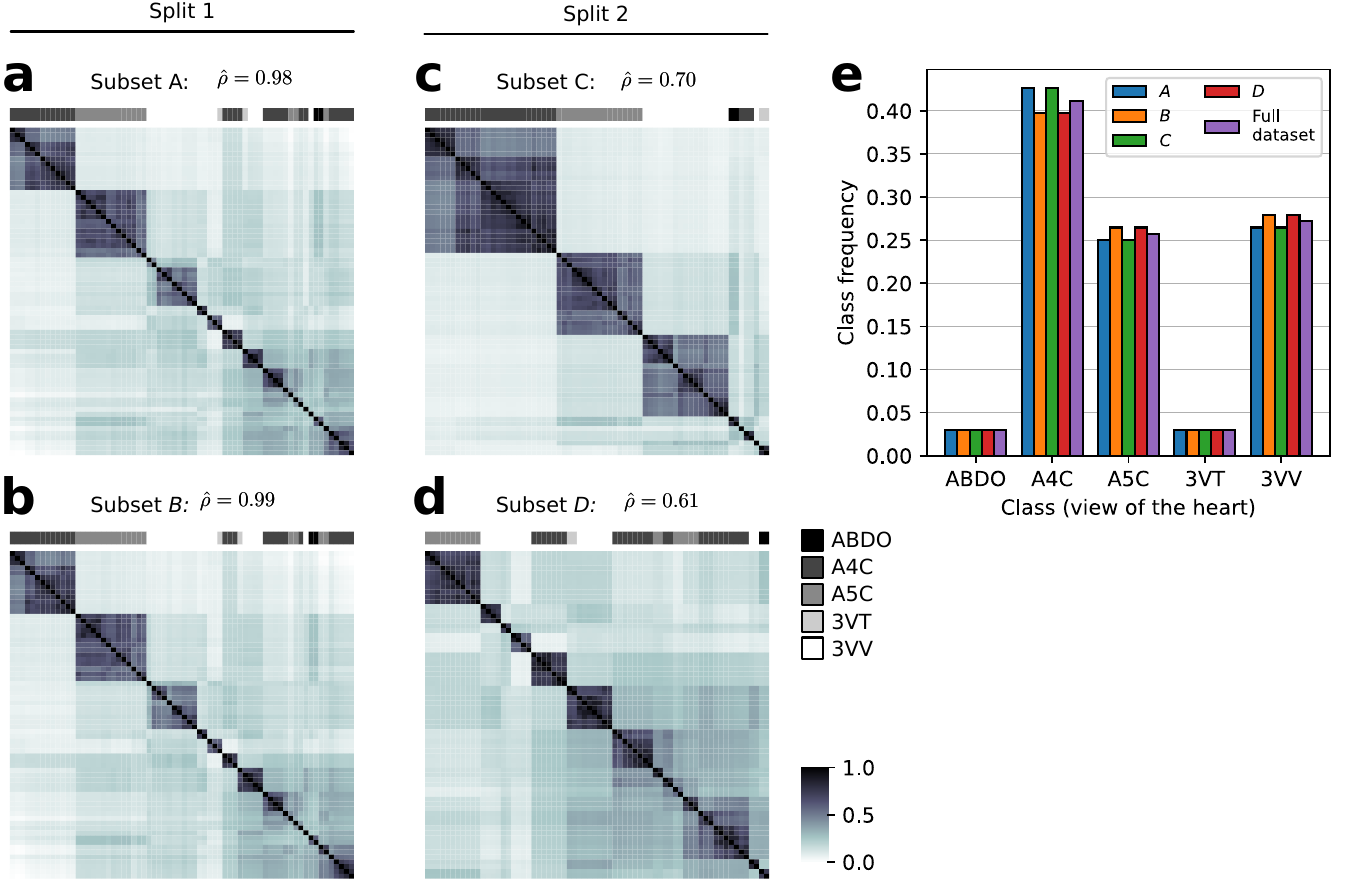}
\caption{
Medical imaging: Representative and non-representative dataset splits. Heatmaps of similarity matrices for subsets of the dataset in Fig. 6. The dataset was split in half into subsets $A$ (a) and $B$ (b), and separately into subsets $C$ (c) and $D$ (d). Although all subsets were identically sized and indistinguishable in terms of class balance (e), $\hat{\rho}_k$ indicates that $A$ and $B$ are redundant while $C$ and $D$ are complementary, reflecting the similarity in clustering pattern in (a) vs. (b) and the difference in (c) vs. (d).
}
\label{fig:7}
\end{figure*}

To illustrate the utility of $\hat{\rho}_k$, consider two different splits of the dataset, both even. Call the subsets that result from the first split A and B and the subsets from the second split C and D. In each case, we assume one of the halves is already labeled. For each split, we calculate $\hat{\rho}_k$ for each subset relative to the whole dataset, using an RMSE-based similarity (Fig. \ref{fig:6}). We find that $\hat{\rho}_A$ and $\hat{\rho}_B$ are both $\sim1$: both are as diverse as the overall dataset, even though they are each only half the size (Fig. \ref{fig:7}). If either of these subsets is labeled, labeling the other adds very little diversity. In contrast, $\hat{\rho}_C$ and $\hat{\rho}_D$ are only 0.6-0.7: labeling the other half would increase diversity substantially.

Currently, the most commonly used heuristics for dataset composition are size and class balance (measured as e.g. the Shannon entropy of the label frequencies). That $A$, $B$, $C$, and $D$ are the same size ($n=68$ images) and have essentially the same class balance (1.27-1.28, vs. 1.28 for the full dataset) yet differ markedly in $\hat{\rho}_k$ demonstrates that this new measure can captures information about datasets that size and class balance alone do not. This ability may be useful in other datasets \cite{37}, \cite{38}, \cite{39} and as a new way to quantify the value of different forms of data augmentation \cite{40}.

\bigskip
\subsubsection{Microscopy/Computational Pathology}

\begin{figure*}[t]
\centering
\includegraphics[width=\textwidth]{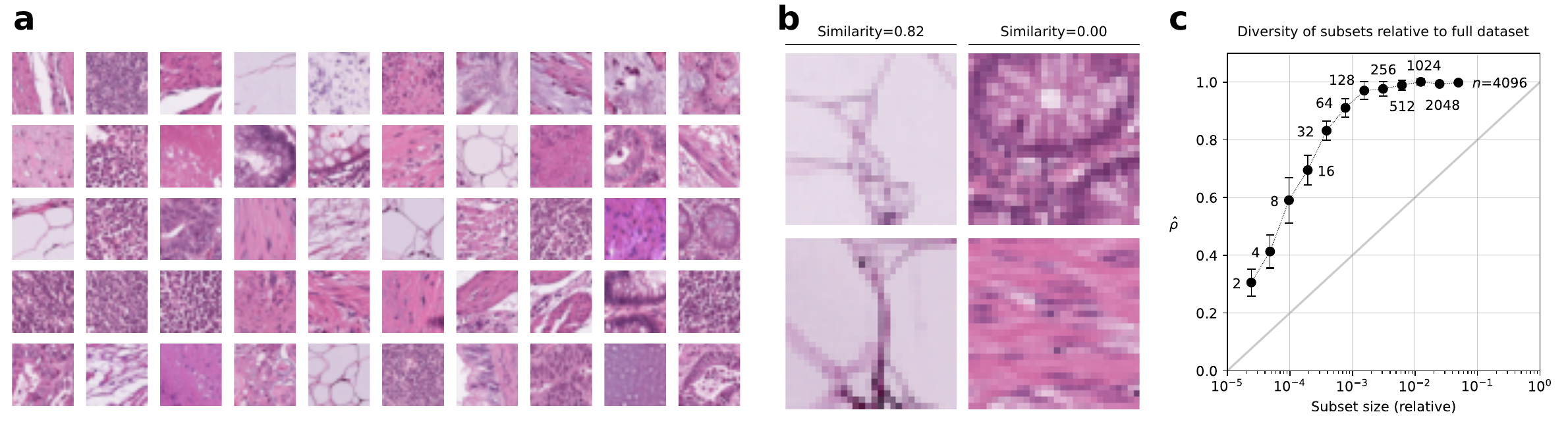}
\caption{
Computational pathology: Capturing dataset diversity in random subsets. (a) Representative selection of images in the PathMNIST training set. (b) Pairs of high-similarity (left) and low-similarity (right) images according to an expert-defined similarity function (see Methods). For comparison, two human subject-matter experts rated the similarity of the left pair as 0.8 and 0.9, and the right pair 0.0 and 0.0. (c) $\bar{\rho}$ for random subsets of images from the dataset. The x-axis indicates what percent of the full dataset each subset is; number of images is listed next to each datapoint. Error bars are s.d. for 10 independent random samples. Diagonal is the 1:1 line. Dotted line is a guide to the eye. The inflection point at 128 images indicates diminishing returns.
}
\label{fig:8}
\end{figure*}

Finally, we illustrate \textit{sentropy} on a much larger ML imaging dataset of histological slide images (computational pathology). We use 89,996 training-set images of haematoxylin/eosin (H\&E)-stained colon tissue sections from the PathMNIST subset of the MedMNIST database \cite{41}. These are $28\times28$-pixel images of patches of 1,000-fold-magnified microscopic fields from each slide that can contain any of the many structures found in healthy and diseased colon tissue, including fat, muscle, and blood vessels as well as colonic epithelium (Fig. \ref{fig:8}a).

Continuing the theme of the previous section, we use \textit{sentropy} to ask how much of the dataset's diversity is captured by randomly chosen smaller subsets. Specifically, we created subsets of increasing size and measure $\bar{\rho}_k$ on each of them, relative to the entire dataset. Because this dataset is much larger than the previous examples, \textit{sentropy}'s on-the-fly similarity matrix option is used, as opposed to pre-calculating the similarity matrix and then loading it from memory as in the previous sections. This is done by passing sentropy a function for calculating the similarities between pairs of images $s_{ij}$. For this example, the function was designed to approximate human pathologists’ determination of pairwise similarity between images (Fig. \ref{fig:8}b), reaching agreement of $R^2=0.48$ with each of two experts (on 100 image pairs) vs. an inter-expert $R^2$ of 0.57. We find that even small subsets capture a substantial fraction of the diversity of the full dataset: for example, subsets of 512 randomly chosen images—less than 1\% of the full dataset—capture $\geq95$\% of the diversity ($p=0.99$) (Fig. \ref{fig:8}c).

\section{Discussion}

Frequency- and especially similarity-sensitive diversity (S-entropy) measures are useful for characterizing complex datasets. Here we have introduced and described \textit{sentropy}, a Python package for calculating them, and demonstrated various applications using examples in immunomics, metagenomics, medical imaging, and computational pathology. \textit{sentropy} is available via PyPI and installed using the standard package manager (pip) for ease of use.

The recasting of simple counting (richness), Shannon entropy, Simpson’s index, and other measures as special cases of a single framework will be new to some investigators but has two advantages. First, it renders these otherwise disparate measures in the same unit—effective number of unique elements, allowing them to be directly compared. And second, it demonstrates that the only difference among them is in how they weight elements' frequencies. However, it recasts investigators' decision of which measure(s) to use as a decision of how much weighting is appropriate for a given investigation: i.e., which q to use. One can avoid this choice by essentially using all of them (e.g. Fig. \ref{fig:4}c-e). This option captures much more information and avoids the possibility of inadvertently cherry-picking a measure that might miss some important feature of the dataset at hand. \textit{sentropy} outputs results for $q=0$ to $\infty$.

In contrast, a choice that cannot be avoided is the choice of similarity measure $s_{ij}$ when calculating S-entropy measures. The same set of elements may be similar in different ways: physiologically, genetically, functionally, constructively, etc. This includes settings that can in principle be quite complex, for example when the similarity between a pair of elements is a function of elements' frequencies, leading to complicated expressions for $s_{ij}$. Because calculation of S-entropy measures ($D_q^Z$) requires $s_{ij}$ for all pairs of n unique elements in the dataset, the time required calculating S-entropy measures scales as $\mathcal{O}(n^2)$. \textit{sentropy} is implemented with parallelization to make use of multicore processors, allowing datasets of $10^5$-$10^6$ elements in under a day on a standard workstation. However, there is opportunity on this rate-limiting step for algorithmic improvement.

The examples presented highlight the complementarity of traditional (similarity-insensitive) and S-entropic (similarity-sensitive) measures (immunomics), the utility of the continuous nature of S-entropy measures (metagenomics), and the value of S-entropy for dataset assembly and assessing dataset quality (machine learning). In the latter context, diversity may prove useful for dataset curation, the need for which has been recognized as a national priority [42]. Given the diversity of these examples, we expect other fields (Table \ref{table:1}) will also benefit from the $D_q^Z$-number Hill framework as brought to Python via the \textit{sentropy} package.

\end{document}